\documentclass[runningheads]{llncs}
\usepackage[T1]{fontenc}
\usepackage{graphicx}
\usepackage[hidelinks]{hyperref} 

\usepackage{pgfplots}
\pgfplotsset{compat=1.18}

\definecolor{colorNever}{RGB}{202,0,32}      
\definecolor{colorRarely}{RGB}{244,165,130}  
\definecolor{colorSometimes}{RGB}{221,221,221} 
\definecolor{colorOften}{RGB}{146,197,222}   
\definecolor{colorAlways}{RGB}{5,113,176}    

\begin{document}
\title{AI Literacy, Safety Awareness, and STEM Career Aspirations of Australian Secondary Students: Evaluating the Impact of Workshop Interventions}
\titlerunning{Evaluating the Impact of AI Literacy Workshop Interventions}

\author{Christian Bergh\inst{1}\orcidID{0009-0000-2424-7356} \and
Alexandra Vassar\inst{1}\orcidID{0000-0001-8856-2566} \and
Natasha Banks\inst{2}\orcidID{0009-0008-8264-1990} \and
Jessica Xu\inst{1,2}\orcidID{0009-0001-2973-2788} \and
Jake Renzella\inst{1}\orcidID{0000-0002-9587-1196}
}
\authorrunning{C. Bergh et al.}
%
\institute{University of New South Wales, Sydney \and
Day of AI Australia
}

%
%

\maketitle              
\begin{abstract}
Deepfakes and other forms of synthetic media pose growing safety risks for adolescents, yet evidence on students' exposure and related behaviours remains limited. This study evaluates the impact of \textit{Day of AI Australia}'s workshop-based intervention designed to improve AI literacy and conceptual understanding among Australian secondary students (Years 7--10). Using a mixed-methods approach with pre- and post-intervention surveys ($N=205$ pre; $n=163$ post), we analyse changes in students' ability to identify AI in everyday tools, their understanding of AI ethics, training, and safety, and their interest in STEM-related careers.

Baseline data revealed notable synthetic media risks: 82.4\% of students reported having seen deepfakes, 18.5\% reported sharing them, and 7.3\% reported creating them.

Results show higher self-reported AI knowledge and confidence after the intervention, alongside improved recognition of AI in widely used platforms such as Netflix, Spotify, and TikTok. This pattern suggests a shift from seeing these tools as merely ``algorithm-based'' to recognising them as AI-driven systems. Students also reported increased interest in STEM careers post-workshop; however, effect sizes were small, indicating that sustained approaches beyond one-off workshops may be needed to influence longer-term aspirations. Overall, the findings support scalable AI literacy programs that pair foundational AI concepts with an explicit emphasis on synthetic media safety.

\end{abstract}
\section{Introduction}
The rapid integration of Artificial Intelligence (AI) into daily life has prompted UNESCO to emphasise the need to prepare learners for AI-mediated futures \cite{Pedro2019ArtificialDevelopment}. In this report, Pedro et al. \cite{Pedro2019ArtificialDevelopment} highlight opportunities of AI-enabled education to improve equity, while stressing the need to strengthen learner capacities. Suggesting that schools must understand existing student engagement with AI and adjust curricula to ensure foundational AI literacy and future preparedness.

Understanding how students perceive AI is therefore essential for designing context-relevant AI literacy initiatives. This study focuses on Years 7--10 students (ages 12--17) from three New South Wales (NSW) government high schools in Australia, who completed pre- and post-intervention surveys as part of the Day of AI Australia curriculum. The surveys provide a view of learners' baseline AI knowledge, confidence using AI tools, ethical understanding (including privacy, bias and deepfakes) and attitudes toward AI's societal impact. Findings provide insight into how secondary school students interpret AI concepts and identify areas for further scaffolded learning, contributing to the growing body of literature on student AI perceptions \cite{Hudec2025StudentEducation,Idroes2023StudentAnalysis,Kim2025MyTasks}.

This paper contributes empirical evidence on the short-term impact of a workshop-based AI literacy intervention implemented in three Australian government secondary schools (Years 7--10; $N=205$ pre, $n=163$ post). We provide (i) a construct-level profile of students' baseline AI literacy, tool-use patterns, and ethical understanding, and (ii) effect size estimates of post-intervention shifts in self-reported AI knowledge, confidence, and key conceptual statements (e.g., training, bias, privacy). We further show how the workshops broadened students' recognition of AI in everyday recommendation systems (e.g., Spotify, Netflix, TikTok), offering a practical lens for designing instruction connecting socio-technical AI concepts to adolescents' lived digital practices. Finally, we highlight the prevalence of deepfake creation and dissemination among participants, and discuss curriculum and policy implications for scaling AI literacy programs in context with heterogeneous teacher confidence and digital infrastructure.

The study is guided by the following research questions:
\begin{enumerate}
    \item[\textbf{RQ1}] How do one-off AI literacy workshops impact students': 
    \begin{enumerate}
        \item perceptions, understandings, misconceptions and confidence of AI? 
        \item aspirations toward Science, Technology, Engineering, and Maths careers?
    \end{enumerate}
    \item [ \textbf{RQ2}] How are Australian government secondary school students engaging with AI tools in their learning and broader digital practices?
\end{enumerate}

\section{Background \& Literature Review}

As AI technologies become increasingly integrated into daily life and work, research emphasises providing structured learning opportunities over incidental exposure \cite{Schiff2021EducationStrategies,Gajos2022DoLearning}. Emerging studies highlight benefits of personalised learning \cite{Jiang2024}, improved feedback \cite{Qin2024}, and accessibility support for diverse learners \cite{Mitre2024UsingDisabilities}. ‘Prompting literacy’, has also been identified as a critical capacity for interacting productively with language models \cite{Xiao2025LearningModule,Gattupalli2023PromptAI}. Studies in primary classrooms demonstrate how teachers have begun co-designing literacy and content lessons that integrate AI resources to support goal-aligned instruction \cite{Kosmas2025IntegratingApproach}.

Current literature has attempted to identify students' mental models and attitudes toward AI, but empirical data on how these perceptions shift following specific curricular interventions remains limited \cite{Marx2023SecondaryReview}. Surveys have indicated that students tend to have an overall positive perception of AI, often citing its potential to increase creativity \cite{Idroes2023StudentAnalysis,Lin2024ArtificialAttitudes}. They also indicate concern for ethics and privacy with AI, but show interest in methods of integrating AI into their learning processes \cite{Idroes2023StudentAnalysis,Smolansky2023EducatorEducation}.

Increasingly, AI literacy initiatives for K--12 students are operationalised through accessible, developmentally suitable pedagogical designs that are adaptable across school systems \cite{Ahmed2024TheAccessibility}. One approach utilises project-based learning to embed AI concepts in creative tasks, such as generating art, music, or text, fostering foundational understanding through direct interaction \cite{Li2024FromEducation}. Another approach addresses equity and infrastructure barriers through “unplugged” curricula that teach AI concepts without digital devices, using structured analogies and games to support reasoning about classification, datasets, and model behaviour in low-resource settings. Unplugged methods yield positive learning gains in conceptual understanding, especially for younger students and schools with limited digital resources \cite{Carrisi2025AEducation,Uema2025DesigningUnplugged}. Beyond these technical foundations, a third pedagogical shift focuses on interdisciplinary socio-technical curricula. This approach moves beyond functional use to engage students with the ethical, civic and societal implications of AI algorithmic automation for communities and institutions \cite{Ratner2025EXPLORINGCOLLABORATION}.
However, despite these emerging global frameworks, empirical data on how these interventions influence the literacy and career aspirations of secondary students, particularly within the Australian context, is limited.

Australian AI education research is nascent, primarily focusing on policy, teacher perspectives or general digital access rather than student perspectives and practices \cite{Blundell2025AdoptingTheory,Creely2025TeachingModel}. Consequently, there is limited evidence to guide system-level decision-making \cite{Elkhodr2024AEducation} or address how local students perceive and interact with AI \cite{Bower2025WhatWhy}. The study utilises workshops provided through the Day of AI Australia program, an initiative providing free curriculum-aligned resources to schools on machine learning, AI ethics, and natural language processing. Although educators value these accessible materials, scaling literacy programs across diverse settings is challenging with variable teacher confidence and digital infrastructure \cite{Renzella2025BuildingClassrooms}. 

School-based AI literacy implementation remains fragmented \cite{Yim2024}, with effectiveness often contingent on teacher confidence, availability, quality of professional learning, and the alignment with local curricula \cite{Scantamburlo2023}. Systemic barriers are further compounded by ethical concerns such as algorithmic bias, data privacy, and the opacity of generative models, potentially undermining educational foundations.  \cite{Alali2024OpportunitiesEducation}. Most existing evidence is from the United States and Europe, yet Australian implementation faces unique hurdles regarding digital equity. Uneven distribution of digital infrastructure, particularly limited internet access, and teacher expertise risk exacerbating existing educational inequalities \cite{Kucuk2020,Bozic2023ArtificalDivide}.Consequently, classroom integration necessitates addressing both infrastructure gaps and the risks associated with data governance and transparency.

\section{Method}
To evaluate differences between pre- and post-workshop samples, we administered pre- and post-intervention surveys consisting of Likert-scale items, multiple-choice questions, and open-ended reflections. The instrument was designed to capture baseline AI literacy and measure shifts in conceptual understanding and career intent. The survey constructs and their respective measurement scales are summarised in \autoref{tab:survey_instrument}. Because surveys were collected anonymously on paper, pre- and post-intervention responses could not be paired at the individual level; therefore, results should be interpreted as differences between pre- and post-workshop samples rather than within-student change. The Day of AI Australia program introduces students to foundational concepts in AI through interactive lessons. Each lesson is described in \autoref{tab:Lessons}.

\begin{table}[t]
\caption{Summary of survey constructs, measurement scales, and item focus.}\label{tab:survey_instrument}
\begin{tabular}{|p{2.5cm}|p{3.4cm}|p{6cm}|}
    \hline
    Construct & \raggedright Measurement Scale (Anchors) & Measurement Focus\\
    \hline
    Demographics & Categorical &  Year level, gender, cultural identity.\\
    AI Knowledge & \raggedright 5pt Likert (None-Expert) & Subjective assessment of AI foundations.\\
    Conceptual App. & Multi-select & Identifying AI in tools (e.g., Spotify, Netflix).\\
    Understanding AI & \raggedright 5pt Likert (Disagree-Agree) & Self assessment of understanding AI functions.\\
    AI Confidence & \raggedright 5pt Likert (Not at all- Extremely) & Self-efficacy in using/troubleshooting AI.\\
    Tool Usage & \raggedright 5pt Frequency (Never-Daily) & Engagement across 12 general and education apps.\\
    Using AI & Multi-select & Identifying primary usage motivations and information sources for AI Literacy.\\
    Deepfakes* & Multi-select & Identifying experience and understanding of deepfakes.\\
    AI Attitudes & \raggedright 5pt Likert (Disagree-Agree) & Ethics, privacy, bias, and societal impact.\\
    STEM Careers & \raggedright 5pt Likert (Uninterested-Interested) & Career intent in AI, CS, and Engineering.\\
    Qualitative (Post-Survey) & \raggedright Open-ended response & Reflections on utility and improvements.\\
    \hline
\end{tabular}
    \scriptsize *Note: Deepfake was defined as: "Video, image or audio generated by, or altered with AI to make it appear as though someone said or did something they did not".
\end{table}

\begin{table}[t]
    \centering
    \caption{Summary of lessons and concepts covered.}
    \begin{tabular}{|c|l|p{7.5cm}|}
    \hline
        \#& Lesson & Description\\
    \hline
         1& What is AI \& How it Works  & Experiment with AI models; discuss how AI learns from data, makes predictions\\
         2& Ethics of AI                & Examine the ethical implications of AI by analysing media examples, biases, and misinformation\\
         3& AI Safety \& Data Privacy   & Examine platform practices, ethical risks of misinformation and deepfakes, and the value of personal data\\
         4& AI \& Misinformation        & Evaluate and distinguish AI-generated content, understand misinformation and disinformation\\
         5& AI in Careers \& Industries & Explore how AI impacts different industries by analysing stakeholders, benefits and risks\\
    \hline
    \end{tabular}

    \label{tab:Lessons}
\end{table}
Participants were recruited from three NSW government secondary schools, across metropolitan and regional contexts ($N = 205$), with a mix of students from lower to higher socio-economic backgrounds. The cohort was culturally diverse, including significant Indigenous representation (17\% at one of the schools) and students who came from language backgrounds other than English (90\% at one school). In accordance with the Ethics approval granted by the Day of AI Australia Human Research and Ethics Committee, participation in all components of the study was voluntary and anonymous. The student cohort was made up of 66 Year 7 students, 69 Year 8 students, 21 Year 9 students, and 48 Year 10 students, and one student did not provide their year. Due to the inclusion of girls-only schools, the sample was majority female ($n = 136$), followed by male ($n = 56$), and students who did not wish to share their gender ($n = 13$). As the senior cohorts (Years 9--10) lacked male representation, data was analysed by total distribution rather than sex to avoid skewed correlations.

\subsection{Procedure}\label{procedure}
The workshop intervention was delivered in a single-day format following a linear progression: 
\textbf{Pre-survey} $\rightarrow$ \textbf{Lessons 1--2} (Foundations \& Ethics) $\rightarrow$ [\textbf{Lessons 3--4} (Safety \& Literacy) \textit{OR} \textbf{Lesson 5} (Careers)] $\rightarrow$ \textbf{Post-survey}. 
The variation in sequence was determined by school-specific scheduling. All survey instruments were administered and collected on-site by program staff for data integrity.

Data was cleaned before analysis. Participants who exhibited straight-lining behaviour (providing identical responses across all items, indicating surveys were not completed in good faith, potentially skewing analyses) were excluded from the final analysis. As the surveys were paper based, it was possible to mark more than one answer or leave questions blank. Responses that were left blank or had multiple selections were deemed invalid and excluded from analysis. Qualitative responses that were illegible were excluded. Written responses that were non-responses such as "none" were excluded. 

Quantitative data were analysed using IBM SPSS Statistics (v30.0.0.0). Due to the non-normal distribution of the Likert-scale data, non-parametric tests specifically the Mann-Whitney U Test and Friedman Test were employed to evaluate pre- and post- intervention shifts. Once the Mann-Whitney U Test for Likert questions present in both pre- and post- surveys were completed, $r$ was calculated for effect size as $r=\frac{z}{\sqrt{N}}$. After conducting the Friedman Test a stepwise post-hoc analysis was conducted on the question asking how much time each student spends on each application. Each application was assigned to groups based on their stepwise subsets. A Wald Independent-Samples Proportions Test was conducted to identify any increase in whether or not they believe a set of applications uses AI.

Qualitative data was analysed using NVivo (v12.0.0.71). A thematic analysis was completed for three open ended survey questions; the first asks what students liked, the second asks what students found most useful, and the third asks what could be done to improve the lessons \cite{Braun2012ThematicAnalysis.}. NVivo's Auto Code wizard was used to identify initial themes and codes. NVivo identified 27 codes for the first question, 18 for the second, and 24 for the third. Codes were grouped into high level themes and then went through three rounds of review for each question and were re-grouped, modified or eliminated each round. After review there were four themes for the first question responses, three for the second, and four for the third.  

\section{Results}

\subsection{Pre-Survey}
Prior to the start of the lessons, $N=205$ student responses were collected (four responses were excluded due to uniform non-authentic answering patterns).

\begin{figure}[t]
\resizebox{\textwidth}{!}{%
\definecolor{colorNever}{RGB}{202,0,32}
\definecolor{colorRarely}{RGB}{244,165,130}
\definecolor{colorSometimes}{RGB}{221,221,221}
\definecolor{colorOften}{RGB}{146,197,222}
\definecolor{colorAlways}{RGB}{5,113,176}

\begin{tikzpicture}
\begin{axis}[
    xbar stacked,
    xmin=-100, xmax=100,
    width=12cm, 
    height=8cm,
    ytick={0,1,2,3,4,5,6,7,8,9,10}, 
    yticklabels={
        Replika,
        EduChat,
        Khanmigo,
        Photomath,
        Grammarly,
        Image Generators,
        Copilot,
        Canva,
        ChatGPT,
        Gemini,
        Social media
    },
    yticklabel style={font=\scriptsize, align=right}, 
    xticklabel style={font=\footnotesize},
    xticklabel={\pgfmathprintnumber\tick\%},
    legend style={
        at={(0.5,-0.12)},
        anchor=north,
        legend columns=-1,
        draw=none,
        cells={align=center},
        /tikz/every even column/.append style={column sep=0.3cm},
        font=\scriptsize
    },
    axis x line*=bottom,
    axis y line*=left,
    xmajorgrids=true,
    grid style={dotted, gray!50},
    bar width=10pt,
    enlarge y limits=0.08,
]

\addplot [fill=colorSometimes, draw=none, forget plot] coordinates {
    (-1.5,0) (-1.6,1) (-1.3,2) (-1.8,3) (-2.3,4) (-6.9,5) 
    (-6.7,6) (-6.3,7) (-7.1,8) (-8.2,9) (-3.9,10)
};
\addplot [fill=colorRarely, draw=none, forget plot] coordinates {
    (-5.7,0) (-3.6,1) (-13.4,2) (-8.3,3) (-14.9,4) (-19.4,5) 
    (-18.6,6) (-28.8,7) (-25.9,8) (-13.8,9) (-12.9,10)
};
\addplot [fill=colorNever, draw=none, forget plot] coordinates {
    (-90.7,0) (-92.7,1) (-79.9,2) (-83.4,3) (-72.2,4) (-55.6,5) 
    (-53.1,6) (-35.1,7) (-31.5,8) (-29.1,9) (-19.1,10)
};

\addplot [fill=colorSometimes, draw=none, forget plot] coordinates {
    (1.5,0) (1.6,1) (1.3,2) (1.8,3) (2.3,4) (6.9,5) 
    (6.7,6) (6.3,7) (7.1,8) (8.2,9) (3.9,10)
};
\addplot [fill=colorOften, draw=none, forget plot] coordinates {
    (0.0,0) (0.5,1) (3.1,2) (4.1,3) (6.2,4) (7.1,5) 
    (11.3,6) (17.3,7) (19.3,8) (28.6,9) (10.3,10)
};
\addplot [fill=colorAlways, draw=none, forget plot] coordinates {
    (0.5,0) (0.0,1) (1.0,2) (0.5,3) (2.1,4) (4.1,5) 
    (3.6,6) (6.3,7) (9.1,8) (12.2,9) (50.0,10)
};

\addlegendimage{area legend, fill=colorNever}
\addlegendentry{Rarely or\\Never}

\addlegendimage{area legend, fill=colorRarely}
\addlegendentry{A few times\\a month}

\addlegendimage{area legend, fill=colorSometimes}
\addlegendentry{Once\\a week}

\addlegendimage{area legend, fill=colorOften}
\addlegendentry{A few times\\a week}

\addlegendimage{area legend, fill=colorAlways}
\addlegendentry{Daily}

\draw [dashed, ultra thin] (axis cs:0,\pgfkeysvalueof{/pgfplots/ymin}) -- (axis cs:0,\pgfkeysvalueof{/pgfplots/ymax});

\end{axis}
\end{tikzpicture}
    }
    \caption{AI tool usage reported by students $N$=197 (191 fully completed).}
    \label{fig:toolusagebar}
\end{figure}
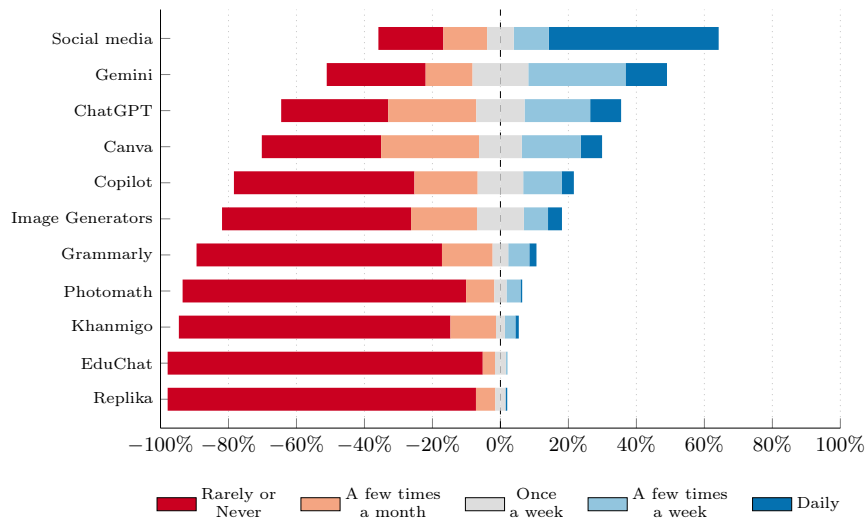

Students were asked to indicate the frequency with which they use specific AI-enabled applications, ranging from 'Rarely or Never' to 'Daily'. \autoref{fig:toolusagebar} presents the distribution of usage across the 11 surveyed applications. Results reveal a clear hierarchy of engagement. AI integrated into social media platforms (Snapchat, TikTok, YouTube) is most used, with most students engaging daily or weekly basis. Generative productivity tools such as ChatGPT, Canva (Magic Write/Design), and Google Gemini show moderate usage, typically weekly use, while dedicated educational AI tools (NSWEduChat and Khanmigo) and companion bots (Replika) are 'Rarely or Never' used by students.

To validate observations, a Friedman Test was conducted to determine differences in usage, returning a significant result ($Q^2 = 667.724, p < .001$). A stepwise post-hoc analysis confirmed five homogeneous subsets, statistically distinguishing the high-frequency social media usage from general generative tools, which were in turn used significantly more often than specialised educational platforms.

Students were asked to indicate their motivations, if any, for using AI tools. As shown in \autoref{fig:ai-motivations}, academic utility is the primary driver, with 'Schoolwork' reported by 65.9\% of students. This was followed by 'Problem Solving' (39.5\%) and 'Fun' (38.0\%), suggesting that while students view AI as a utility for learning, recreational and creative exploration remain significant factors.

When students seek out information about AI, they indicated a heavy reliance on social media (74.1\%), teachers (35.6\%) and friends (34.6\%) (\autoref{fig:ai-sources}). Additionally, students were asked if they learned about AI at school, with 42.1\% reporting 'Yes', 28.4\% reporting 'No', and 29.5\% reporting 'Unsure'.

\begin{figure}[!b]
    \centering
    \begin{minipage}[t]{0.48\textwidth}
        \centering
        \resizebox{\linewidth}{!}{%
            \begin{tikzpicture}
\begin{axis}[
    xbar, 
    width=\linewidth, 
    height=7cm,       
    xmin=0, xmax=80,  
    ytick={0,1,2,3,4,5,6,7}, 
    yticklabels={
        Friendship,
        Other,
        Communication,
        Advice,
        Creativity,
        Fun,
        Problem Solving,
        Schoolwork
    },
    yticklabel style={
        font=\small,
        align=right,
        text width=2.5cm, 
        anchor=east
    },
    xticklabel={\pgfmathprintnumber\tick\%},
    nodes near coords={%
        \pgfmathprintnumber[fixed, precision=1]\pgfplotspointmeta\%
    },
    nodes near coords style={
        font=\footnotesize, 
        anchor=west, 
        black
    },
    axis x line*=bottom,
    axis y line*=left,
    xmajorgrids=true,
    grid style={dotted, gray!50},
    enlarge y limits=0.1,
    bar width=14pt,
]

\addplot[
    fill={rgb:red,5;green,113;blue,176}, 
    draw=none
] coordinates {
    (10.7,0)  
    (14.1,1)  
    (14.6,2)  
    (22.9,3)  
    (30.7,4)  
    (38.0,5)  
    (39.5,6)  
    (65.9,7)  
};

\end{axis}
\end{tikzpicture}
        }
        \caption{Primary motivations for AI tool engagement among students.}
        \label{fig:ai-motivations}
    \end{minipage}
    \hfill 
    \begin{minipage}[t]{0.48\textwidth}
        \centering
        \resizebox{\linewidth}{!}{%
            \begin{tikzpicture}
\begin{axis}[
    xbar, 
    width=\linewidth, 
    height=6cm,       
    xmin=0, xmax=85,  
    ytick={0,1,2,3,4,5,6}, 
    yticklabels={
        Other,
        Parents,
        I Don't look,
        AI Tools,
        Friends,
        Teacher,
        Social Media
    },
    yticklabel style={
        font=\small,
        align=right,
        text width=2.5cm, 
        anchor=east
    },
    xticklabel={\pgfmathprintnumber\tick\%},
    nodes near coords={%
        \pgfmathprintnumber[fixed, precision=1]\pgfplotspointmeta\%
    },
    nodes near coords style={
        font=\footnotesize, 
        anchor=west, 
        black
    },
    axis x line*=bottom,
    axis y line*=left,
    xmajorgrids=true,
    grid style={dotted, gray!50},
    enlarge y limits=0.1,
    bar width=14pt,
]

\addplot[
    fill={rgb:red,5;green,113;blue,176}, 
    draw=none
] coordinates {
    (5.9,0)   
    (22.0,1)  
    (28.3,2)  
    (29.8,3)  
    (34.6,4)  
    (35.6,5)  
    (74.1,6)  
};

\end{axis}
\end{tikzpicture}
        }
        \caption{Primary information sources for student AI literacy.}
        \label{fig:ai-sources} 
    \end{minipage}%
   
\end{figure}

Responses regarding deepfakes indicate high exposure. Most students (82.4\%) reported having seen deepfakes, and some reported sharing (18.5\%) or creating (7.3\%) them. Some (13.7\%) reported seeing their friends create or share deepfakes. Crucially, gaps in safety knowledge exist; while half of the cohort (50.7\%) indicated they knew how to report deepfakes and 14.1\% having reported a deepfake, only 34.1\% expressed confidence in knowing what to do if they were the subject of one and 6.8\% indicated they have been a subject of a deepfake.

\subsection{Post-survey}
Post-workshops, $N=163$ student responses were collected (seven responses were excluded due to survey straight-lining), and 35 students did not submit a post-workshop survey.

A primary objective of the intervention was to broaden student definitions of AI beyond generative text/image tools. As detailed in \autoref{tab:DoToolsUseAI}, independent samples proportions tests revealed significant shifts in students' ability to identify AI within algorithmic recommendation engines. While recognition of overt AI tools like ChatGPT remained high (ceiling effect), significant increases were observed for Netflix ($+27.1\%, p<.001$), TikTok ($+12.8\%, p<.001$), Spotify ($+16.2\%, p<.001$), and Face Unlock ($+11.2\%, p=.013$). This suggests the workshop challenged the misconception that "algorithms" are distinct from AI.

\begin{table}[!b]
        \caption{Proportion of respondents identifying tools as using AI, with one-sided Wald Test results.}
    \label{tab:DoToolsUseAI}
    \begin{tabular}{|p{4cm}|c|c|r|c|c|c|c|}
    
    \hline
    & \multicolumn{2}{|c|}{Proportion}&&&& \multicolumn{2}{|c|}{CI} \\
    &  Pre &Post &\multicolumn{1}{|c|}{\% Change} &$z$ & $p$ &Lower&Upper \\
    \hline
         Calculator & 28.8\% & 30.1\% & 1.3\%  &-.268& .394 &-.107 &.080    \\
         Spotify    & 60.5\% & 76.7\% & 16.2\% &-3.300& \textbf{<.001} & -.253 & -.067 \\
         TikTok     & 82.9\% & 95.7\% & 12.8\% &-3.830& \textbf{<.001} & -.187 & -.064  \\
         Microwave  & 10.2\% & 11.0\% & .8\%  &-.247& .402    & -.073 & .055\\
         Netflix    & 38.5\% & 65.6\% & 27.1\% &-5.166& \textbf{<.001} & -.366& -.170  \\
         ChatGPT    & 94.1\% & 93.3\% & -.9\%  &.352& .362    & -.042& .062 \\
         Google     & 86.3\% & 93.3\% & 6.9\%  &-2.139& \textbf{.016} &-.129 & -.006   \\
         Face Unlock& 60.0\% & 71.2\% & 11.2\% &-2.230& \textbf{.013} &-.206 & -.014   \\
    \hline
    \end{tabular}
    \smallskip
\begin{minipage}{\linewidth}
    \scriptsize Note: For each proportion $N=368= (nPre=205+nPost=163)$. Significance level: .05
\end{minipage}
\end{table}


Mann-Whitney U Tests were conducted to evaluate shifts in self-reported knowledge and attitudes (\autoref{tab:AiKnowledgeUTest}). The intervention yielded the largest effect sizes in conceptual understanding. Students reported a significant increase in general 'AI Knowledge' ($r=.445$) and 'AI Confidence' ($r=.279$) with a medium and small effect size respectively. Additionally, medium effect sizes were also observed regarding their understanding of how AI systems are trained ($r=.399$) and their understanding of bias ($r=.302$). Alternatively, the intervention did not appear to impact the students' perception of AI, the only small effect change occurring in students' worry of AI's impact on job markets ($p=.049,r=.106$). Regarding career aspirations, the data indicates an uplift in interest across Artificial Intelligence ($p<.001$), Computer Science ($p=.005$), and Engineering ($p=.029$). However, the effect sizes for these career variables remained small ($r < .3$), suggesting that while the workshop successfully piqued interest, a single-day intervention may be insufficient to meaningfully alter long-term career trajectories compared to the immediate gains in content knowledge.

\begin{table}[!t]
    \caption{
    Comparison of pre- and post-intervention AI knowledge \& confidence scores, AI knowledge statements, and AI sentiment statements.}
    \begin{tabular}{|p{6.6cm}|c|c|c|c|c|c|}
    \hline
        & \multicolumn{2}{|c|}{$n$}&&&& \\
        Confidence/Statement/Career Interest &	Pre&	Post&	$U$& $p$ &	$z$&	$r$ \\
    \hline
        Knowledge about AI&	200&	158&	7945.0&	\textbf{<.001}&	-8.428&	\textbf{.445}\\ 
        Confidence in AI Tools&	179&	154&	9567.5&	\textbf{<.001}&	-5.084&	.279 \\
                \hline
        Understand AI training &	201&	160&	8902.0&	\textbf{<.001}&	-7.587&	\textbf{.399}\\
        Understand AI bias/errors &	200&	160&	10636.5&\textbf{<.001}&	-5.725&	\textbf{.302} \\
        Privacy skills (Social Media)&	201&	157&	11475.0&	\textbf{<.001}&	-4.637&	.245 \\
        Privacy skills (School Work) &	200&	158&	11997.0&	\textbf{<.001}&	-4.122&	.218 \\
        Importance of data consent&	199&	157&	12617.5&	\textbf{<.001}&	-3.302&	.175 \\
        \hline
        Worried AI impacts my job&	187&	157&	12943.5&	\textbf{.049}&	-1.965&	.106 \\
        AI tools at school to help with learning&	187&	156&	14333.0&	.768&	-.296&	.016\\
        I feel pressured to use AI tools&	184&	155&	12695.5&	.068&	-1.827&	.099\\
        Worried teacher suspects AI&	184&	156&	13933.0&	.628&	-.484&	.026 \\
        Effect of AI on society will be positive&	176&	149&	11973.0&	.090&	-1.695&	.094\\
        \hline
        Career in Artificial Intelligence & 175& 157& 10848.0 & \textbf{<.001} & -3.443& .189\\
        Career in Computer Science        & 176& 157& 11424.5 &\textbf{.005}  & -2.831& .115\\
        Career in Engineering             & 178& 157& 12094.5 &\textbf{.029}  & -2.178& .119\\
    \hline
    \end{tabular}
    \smallskip
\begin{minipage}{\linewidth}
\scriptsize  Note: Statistical significance was determined using the Mann-Whitney U Test ($N=nPre+nPost$).
\end{minipage}
    \label{tab:AiKnowledgeUTest}
\end{table}

Thematic analysis of open-ended responses ($N=133$) reinforced the quantitative findings regarding knowledge acquisition. When asked what was "most useful" 85 (63.9\%) students cited 'How AI works/trained'. Students identified the interactive elements and gamified components ($n=52$, $39.0\%$) as the most engaging aspects of the curriculum. Similarly, suggestions for improvement focused on hands-on activities ($n=60$, $45.1\%$), with students requesting "more activities and less talking," highlighting a preference for kinaesthetic learning over direct instruction in technical workshops.

\section{Discussion}
Students identified social media-integrated AI (Snapchat, TikTok and YouTube) as their most frequently used AI applications. This likely reflects the entrenched role of social media in teenage routines \cite{Cervi2021TikZ,MichelleFaverio2024TeensTechnology} (RQ2). Notably, this survey took place prior to new Australian legislation restricting social media access for those under 16 \cite{ClareArmstrong2026MoreReveals}. As accounts are closed or transitioned, teen engagement patterns may shift toward more task-specific or general AI tools.

Despite 65.9\% of students self-reporting schoolwork as a primary motivation, 57.9\% report that they are unsure or did not learn about AI at school (RQ2). This result highlights a gap in AI literacy within Australian school environments.

While students significantly improved at identifying AI in complex platforms like Spotify, Netflix, and TikTok, distinguishing AI from basic tools like calculators remains a challenge. A general increase in 'Yes' responses suggests potential response bias, yet this cannot explain the disproportional gains seen in identifying AI within social media and streaming platforms (RQ1a). This shift indicates a narrowing terminological disconnect; where these platforms were once marketed as "algorithms", students now increasingly recognise the underlying AI. As AI becomes increasingly invisible within daily-use software, students must be able to identify its presence to navigate these systems critically and ethically. 

Qualitative feedback revealed a strong preference for hands-on activities, with students identifying interactive components as the most enjoyable workshops elements. To maintain student agency and attention, educational AI platforms should consider adopting engagement strategies from general-use applications to make niche educational tools more appealing to adolescents.

While the workshops achieved a statistically significant increase in STEM career interest, the effect size was low (RQ1b). This suggests that while a single-day intervention can spark curiosity, sustained, longitudinal engagement is required to meaningfully shift long-term career trajectories.

\subsection{Implication for Policy}
AI policy frameworks for education, including the \textit{Australian Framework for Generative AI in Schools}, UNESCO’s \textit{Guidance for Generative AI in Education and Research}, and UNICEF’s \textit{Policy Guidance on AI for Children} emphasise the importance of ethical, inclusive, and curriculum-aligned AI literacy education. We find that while one-off workshop interventions can make significant headway in improving student perceptions and literacy, influencing career aspirations likely requires a more integrated, sustained approach. For policy-makers, this differentiation is important as it underscores that whilst these types of workshops are effective entry points, they must be supported by systemic curriculum integrations to drive workforce readiness. Education policy must be grounded in evidence that captures both immediate literacy gains, and long-term outcomes. As AI continues to reshape the global economy and workforce, policy must prioritise equitable access to ensure that the "intelligence gap" does not widen for students in disadvantaged or under-represented cohorts. 

\subsection{Limitations and Future Work}
Due to the challenging nature of research of Australian government secondary schools, this study prioritised accessibility and low-friction participation, which influenced the research design and identified several avenues for future work. This most significancy manifested in the 35 student drop-off between pre- and post- responses from one school (N=136) which could impact some findings of this study, future work should implement measures to mitigate drop-off.

Two of the three schools were all-girls institutions. While efforts were made to include diverse cohorts, some findings may not be generalisable. Future research should explicitly investigate gender-based responsiveness to AI learning materials to determine if specific pedagogical approaches are more effective for different cohorts.

As student anonymity during paper-based data collection was required, pre- and post-intervention surveys could not be paired at the individual level. This necessitated cross-sectional analysis using Mann-Whitney U Test, rather than a longitudinal, paired-samples approach. Future studies could utilise paired methods to provide more granular understanding of impact on individuals.

Due to school-specific scheduling, there was a variation in the specific lessons each cohort completed (\autoref{procedure}), certain lessons may have contributed more significantly to the observed outcomes. Controlled experiments will provide clearer signals for future interventions.

Findings were based on self-reported measures of AI knowledge and confidence rather than objective performance data. Specifically, reported deepfake creation (7.3\%) and sharing (18.5\%) should be interpreted with caution as these figures may be influenced by acquiescence bias. Although several AI literacy constructs are currently in development, a gold standard quantitative tool has yet to emerge \cite{Almatrafi2024A20192023}. Future research could incorporate these frameworks to quantify gains in AI literacy beyond student perception. 

Finally, this study provides a snapshot of immediate impact, no claims can be made regarding the sustained impact or durability of these interventions on long-term knowledge retention, attitudes, skills, or career trajectories. Longitudinal research will better inform school-, state-, and national-level education policy both within Australia and internationally.

This study featured a high proportion of Indigenous participants at 17.1\%, unevenly distributed across the three schools. The study did not isolate their specific perceptions or responsiveness to lessons. Opportunities exists for future work to prioritise Indigenous viewpoints to develop culturally responsive AI frameworks that promote equity and inclusion within the Australian educational landscape.

\section{Conclusion}
This study evaluated the efficacy of the Day of AI Australia's workshop interventions in fostering AI literacy and STEM career aspirations among Australian secondary students. Our findings demonstrate that targeted, short-duration workshops can facilitate statistically significant increases in self-reported AI knowledge ($p < .001, r = .445$) and self-efficacy. Notably, the interventions successfully equipped students to better identify AI within their daily lives. This shift is vital for developing a citizenry capable of navigating AI safely, ethically, and critically.

While the interventions resulted in increased interest toward AI, Computer Science, and Engineering careers ($p < .05$), the relatively small effect sizes suggest that singular interventions may serve more as a catalyst for interest, rather than a definitive driver of long-term career trajectories. This highlights the necessity for AI literacy to be integrated more deeply and systemically into Australian state curricula and the National Curriculum. Future research should focus on longitudinal studies to determine the persistence of literacy gains and whether sustained exposure leads to more robust shifts in long-term STEM participation. Ultimately, as AI reshapes the global labour market, providing equitable access to foundational AI education remains a primary lever for preventing a widening digital and intelligence gap within Australia and internationally.


\begin{credits}
\subsubsection{\ackname} Day of AI Australia and its research activities are funded by philanthropic donations and corporate partnerships.

\subsubsection{\discintname} 
Institution 2 is a registered Australian charity. Institution 1 and 2 have a legal collaboration. Institution 2 has previously received funding from Google.org and Microsoft, contributions have not funded this research.



\end{credits}
%
%
%

\bibliographystyle{splncs04}
\bibliography{sample-base}
\end{document}